# Polymer Crystallization in 25 nm Spheres

Yueh-Lin Loo and Richard A. Register
Department of Chemical Engineering, Princeton University
Princeton, NJ 08544-5263

Anthony J. Ryan
Department of Chemistry, University of Sheffield
Sheffield S3 7HF United Kingdom

*Abstract*

Crystallization within the discrete spheres of a block copolymer mesophase was studied by time-resolved x-ray scattering. The cubic packing of microdomains, established by self-assembly in the melt, is preserved throughout crystallization by strong interblock segregation even though the amorphous matrix block is well above its glass transition temperature. Homogeneous nucleation within each sphere yields isothermal crystallizations which follow first-order kinetics, contrasting with the sigmoidal kinetics normally exhibited in the quiescent crystallization of bulk polymers.

Block copolymers provide well-defined model systems for studying self-assembly, since the molecular architecture and interactions can be precisely varied over a wide range [1,2]. When the repulsive interblock interactions are sufficiently large, molten block copolymers spontaneously self-assemble into mesophases having morphologies such as spheres, cylinders, and lamellae. While these mesophases may be likened to crystals, with a characteristic spacing comparable to the macromolecular size, the most common self-assembly process in nature is simple crystallization, where the characteristic spacing is of atomic dimensions. In polymers, the crystallites which form are typically thin (order 10 nm) in one dimension and of comparatively large lateral extent, leading to micron-scale superstructures such as the familiar polymer spherulites [3]. Recent studies have probed crystallization in ultrathin polymer films on a substrate [4-6], revealing that both the crystallinity and crystallization kinetics can be strongly perturbed by such confinement.

Combining two or more self-organizing mechanisms [7], such as crystallization and interblock repulsion, into a single polymer yields morphological richness and kinetic complexity. For example, when a block copolymer is weakly segregated and both blocks are above their glass transition temperatures, the melt mesophase is generally destroyed when one block crystallizes [8,9]; crystallization creates an entirely new structure bearing little resemblance to that present in the melt. However, by increasing the segregation strength, the melt mesophase may effectively confine crystallization to regions of predefined size, shape, and connectivity. If so, how is the crystallization process perturbed when crystals are forced to be thin in two or three dimensions, instead of just one? If crystallization can be confined to discrete microdomains, does homogeneous

nucleation—normally of little importance in polymer crystallization—dominate? In this Letter, we explore these issues in a diblock copolymer where strong interblock segregation is used to restrict crystallization to spheres of 25 nm diameter.

The diblock studied here, E/SEB63, contains a short polyethylene (E) block connected to a longer styrene-ethylene-butene (70/14/16 by weight) random terpolymer block. The diblock was synthesized by sequential anionic polymerization of butadiene, then of a styrene/butadiene mixture [10], followed by catalytic hydrogenation [11]. The E block, which constitutes 14.3 wt% of the diblock, is itself effectively a random copolymer (containing 8 wt% butene) due to the microstructure of the precursor polybutadiene. The diblock has a weight-average molecular weight $M_w$ of 63 kg/mol, a polydispersity of 1.16, and no unattached E chains. The amorphous SEB matrix has a glass transition temperature of 25$^o$C by differential scanning calorimetry (DSC). E/SEB63 has a final melting temperature of 107$^o$C by DSC and a weight fraction crystallinity of the E block of 0.32 at room temperature. For comparison, an E "homopolymer" (E40) of the same microstructure as the E block in E/SEB63 (8 wt% butene, so E40 is actually a random copolymer) was separately synthesized, with $M_w$ = 40 kg/mol, a polydispersity of 1.05, and a room-temperature crystallinity of 0.30.

Figure 1 shows a room-temperature transmission electron micrograph of a microtomed thin section of E/SEB63, stained with $RuO_4$. E spheres of regular size and spacing are readily apparent, indicating that interblock segregation is indeed sufficient to confine crystallization to 25 nm spheres. This particular specimen was isothermally crystallized at 70$^o$C in a DSC pan, following cooling from the melt. To study the crystallization process in real time, we conducted time-resolved small-angle (SAXS) and wide-angle (WAXS) x-ray scattering measurements on beamline 8.2 at the Daresbury Laboratory. As shown in Figure 2, in the melt, E/SEB63 exhibits an intense narrow primary peak at $q^* = 0.19$ nm$^{-1}$ and higher-order peaks at $q/q^*$ ratios of $\sqrt{2}$ and $\sqrt{3}$, indicating a body-centered cubic (bcc) packing of E spherical microdomains [12] in an SEB matrix. Based on the composition, the known densities of the materials constituting each block, and the primary peak position, the individual spherical microdomains are calculated to have a 12.7 nm radius. The dashed curve in Figure 2 shows the shape of the theoretical scattering [13] from isolated spheres of this radius; the local minima observed in the melt at 0.35 and 0.6 nm$^{-1}$ are clearly the form factor nodes of the individual spheres, confirming the structure. This polymer does not disorder up to its decomposition point (>300°C), indicating strong interblock segregation. Following isothermal crystallization at 66°C (upper curve in Figure 2), the form factor minima and maxima are washed out, replaced by a single broad maximum centered near 0.5 nm$^{-1}$. That the material had reached its limiting crystallinity was confirmed through the simultaneously-acquired WAXS patterns. The narrow structure factor peaks remaining at $q^*$ and $\sqrt{2}q^*$ indicate that the bcc lattice is preserved, exactly as one would expect from the TEM image in Figure 1, and hence that the E blocks crystallize within the self-assembled microdomains rather than forming laterally-extended lamellar crystals. The elimination of clear form factor minima reflects distortions of the microdomains away from perfect sphericity, while the enhanced scattering in this $q$-range results from the additional heterogeneity created by small crystallites within the E spheres.

Isothermal crystallizations of E/SEB63 were followed via the time evolution of the SAXS and WAXS patterns. Figure 3 shows the integrated intensities of the broad SAXS maximum ($q \sim 0.5$ nm$^{-1}$) and the polyethylene (110) WAXS reflection as E/SEB63 is isothermally crystallized at 67°C. The SAXS and WAXS intensity *vs.* time curves start off steeply sloped and asymptotically reach their final values [14]. This contrasts sharply with the conventional sigmoidal kinetics which E40 exhibits. The isothermal crystallization of E40 is well-described by the Avrami equation [15]:

$$\ln(1-X) = -Kt^n \tag{1}$$

where X is the fraction of the transformation which has occurred at time t, and the prefactor K and exponent *n* are referred to as the Avrami constants ($n = 2.4$ for E40 at 95°C). By contrast, the crystallization kinetics for E/SEB63 are quantitatively well described by a simple exponential decay, or $n = 1$ in the Avrami equation. Such first-order kinetics indicate that the rate of isothermal crystallization is simply proportional to the fraction of spheres which have yet to crystallize, as anticipated if crystallization in E/SEB63 is confined within individual microdomains.

Further confirmation that each E microdomain crystallizes in isolation comes from varying the temperature during crystallization. E/SEB63 was first quenched from the melt to a temperature where crystallization is relatively rapid, and held there until an intermediate level of crystallinity developed. The sample was then quickly ramped to a slightly higher temperature where the isothermal crystallization rate is slow. Data from a representative jump experiment on E/SEB63 are shown in Figure 4, with isothermal crystallization traces included for comparison. The data in the jump experiment initially follow the 66°C isothermal data (when the sample is at 66°C), and later follow the 70°C isothermal data (when the sample is at 70°C), exactly as expected if two isolated populations of material are crystallized at the two temperatures. By contrast, if nuclei generated at 66°C could continue to crystallize material after raising the temperature to 70°C, only a modest reduction in crystallization rate (due to the slower crystal growth rate) would be seen after jumping to 70°C.

We expect the E spheres in E/SEB63 to be homogeneously nucleated because the number of microdomains (~ 2 x $10^{16}$ spheres/cm$^3$) far exceeds the possible number of impurities in the sample (of order $10^9$ /cm$^3$, as a semicrystalline homopolymer typically shows 10 µm diameter spherulites). The nucleation process in polymers at the deep undercoolings needed for homogeneous nucleation has been of significant interest recently [16-18]. The idea that homogeneous nucleation could be prevalent in block copolymers with crystallizable minority blocks and a vitreous matrix was put forth originally by Lotz and Kovacs [19], and supported by the work of Robitaille and Prud'homme [20] which showed that very high undercoolings were needed to crystallize the poly(ethylene oxide) minority blocks in a triblock with a vitreous matrix. In our case, though the matrix is well above its glass transition temperature during crystallization, strong interblock segregation is equally effective at confining crystallization [21]. Because the E spheres within E/SEB63 are only 25 nm across, crystal growth from the nucleus to the microdomain interface is essentially instantaneous, and the temperature dependence of the crystallization rate in E/SEB63 reflects the temperature dependence of the nucleation rate only. Figure 5 shows the relationship between crystallization half-time and crystallization temperature, $T_c$, for both E/SEB63 and E40. Relative to the final melting temperatures determined by DSC, E/SEB63 requires a significantly larger undercooling (37°C) than does E40 (7°C) for both materials to crystallize with a half-time of 10 min. These undercoolings are comparable to those found for the homogeneous and heterogeneous nucleation of micron-sized droplets of linear polyethylene (PE) homopolymer suspended in [22,23] or on [24,25] various oils. Figure 5 shows that for E/SEB63, the crystallization time increases by a factor of 2.9 for every 1°C increase in undercooling. This steep temperature dependence is characteristic of homogeneous nucleation and compares adequately with factors of 3.4/°C [24] and 4.7/°C [25] extracted from literature data for the homogeneous rates of PE homopolymer droplets. The weaker temperature dependence of the half-time for E40 (1.4/°C) reflects the combined temperature dependences of heterogeneous nucleation and spherulitic growth, both of which are less temperature-dependent than homogeneous nucleation. Finally, we note that the temperature range where E/SEB63 shows a half time of 1-30 min (66-69°C) is actually about 10°C lower than the temperature range found by Barham *et al.* [25] for crystallization of PE droplets at the same rate. While a detailed comparison is complicated by the minor content of butene units in our E block, which may affect the nucleation rate, this 10°C temperature difference can be entirely explained by the difference in the diameters of the E domains (25 nm here *vs.* approximately 1 μm in [25]), as the half-time is inversely proportional to the droplet volume.

First-order crystallization kinetics ($n = 1$) are highly unusual; polymers typically show Avrami exponents [15] ranging from 2 – 4, reflecting growth in two or three dimensions from isolated nuclei. Polymers stretched at extremely high rates have shown first-order kinetics, presumably because alignment of the chains induces massive nucleation [26,27]. However, our materials are quiescent during crystallization. Unusual Avrami exponents (≈1/2) were also found by Lotz and Kovacs [19] for block copolymers with a glassy matrix and a crystallizable minority block; the extreme smallness of the exponent there may reflect the constraints imposed by the glassy matrix, rather than the crystallization process itself, as the kinetics were followed by dilatometry. Moreover, in this case a measurable fraction of the material crystallized at the shallow undercoolings characteristic of heterogeneous nucleation [19]. Mixed homogeneous and heterogeneous nucleation is typical in droplet experiments as well, even with extensive sample purification [23,25], simply because it is quite difficult to produce submicron droplets. By contrast, the E/SEB63 diblock shows "clean" homogeneous nucleation of the E blocks, and since it is a bulk material, bulk techniques with good time resolution (SAXS, WAXS, DSC) can be used to follow even fast transformation processes. Moreover, the microdomain volume, which must be known to extract a nucleation rate, is essentially monodisperse in the block copolymer case, where it can also be precisely measured and controlled, in contrast to the conventional "droplet" experiments. This approach should be extendable to any system where a block, graft, or segmented copolymer can be formed between units with large repulsive interactions.

Financial support for this study came from the National Science Foundation, Polymers Program (DMR-9711436) and the NATO Collaborative Research Grants Program (CRG 951243), while Daresbury beamtime was provided by the EPSRC. We thank Christine Kliewer of Exxon Corporate Research for instruction in TEM sample preparation.

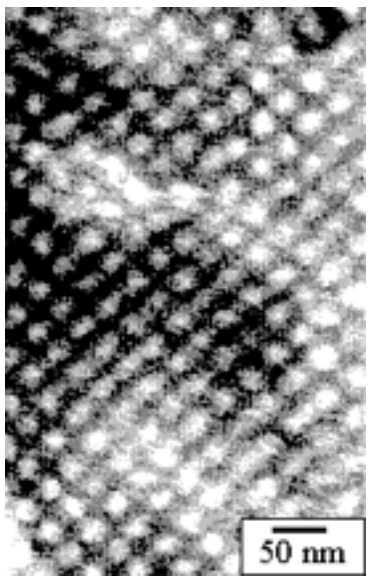

Figure 1. Transmission electron micrograph of a thin section of E/SEB63. E domains appear as light circles with a spacing of 30-40 nm.

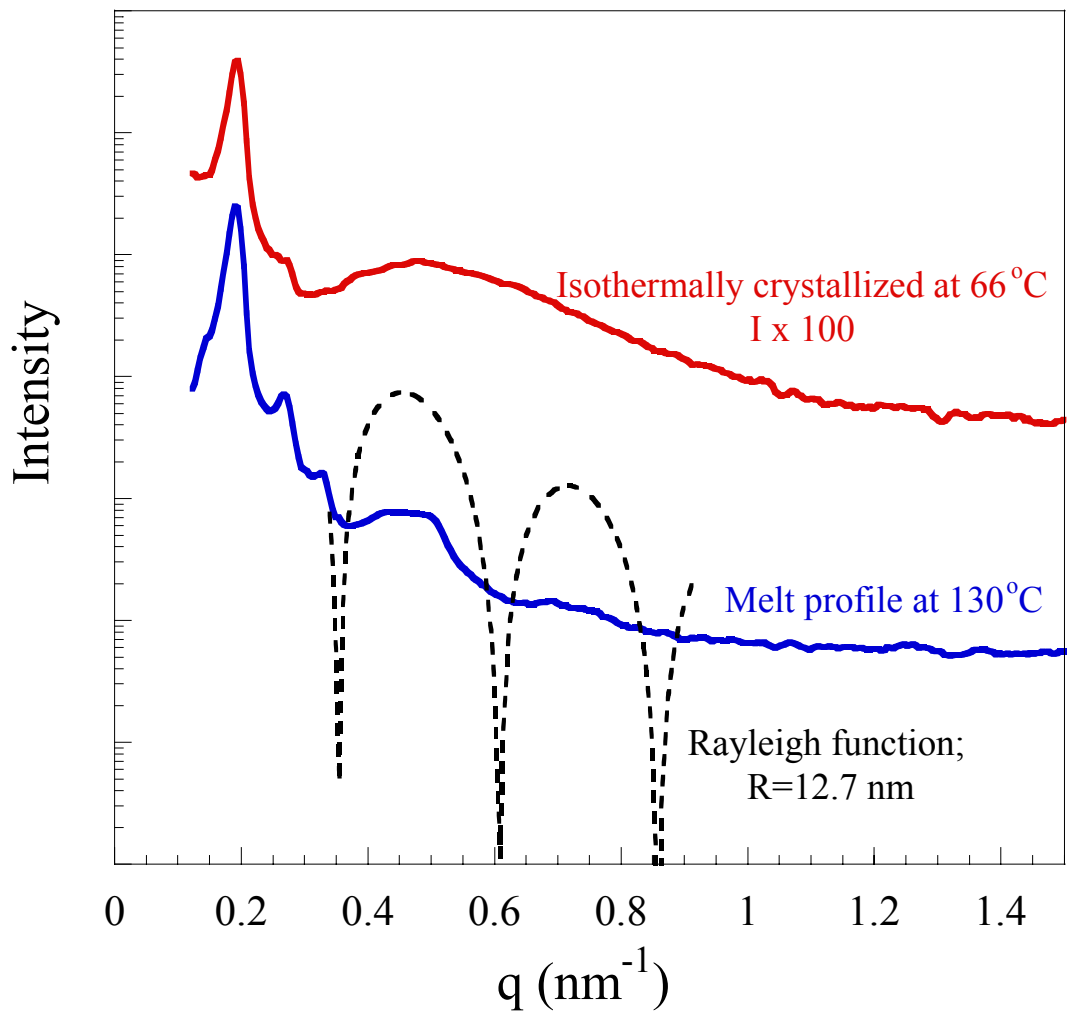

Figure 2. SAXS profiles of E/SEB63 in the melt at 130°C (bottom) and at 66°C after crystallization (top). Dashed curve shows the calculated scattering from isolated spheres of 12.7 nm radius (arbitrary intensity scale).

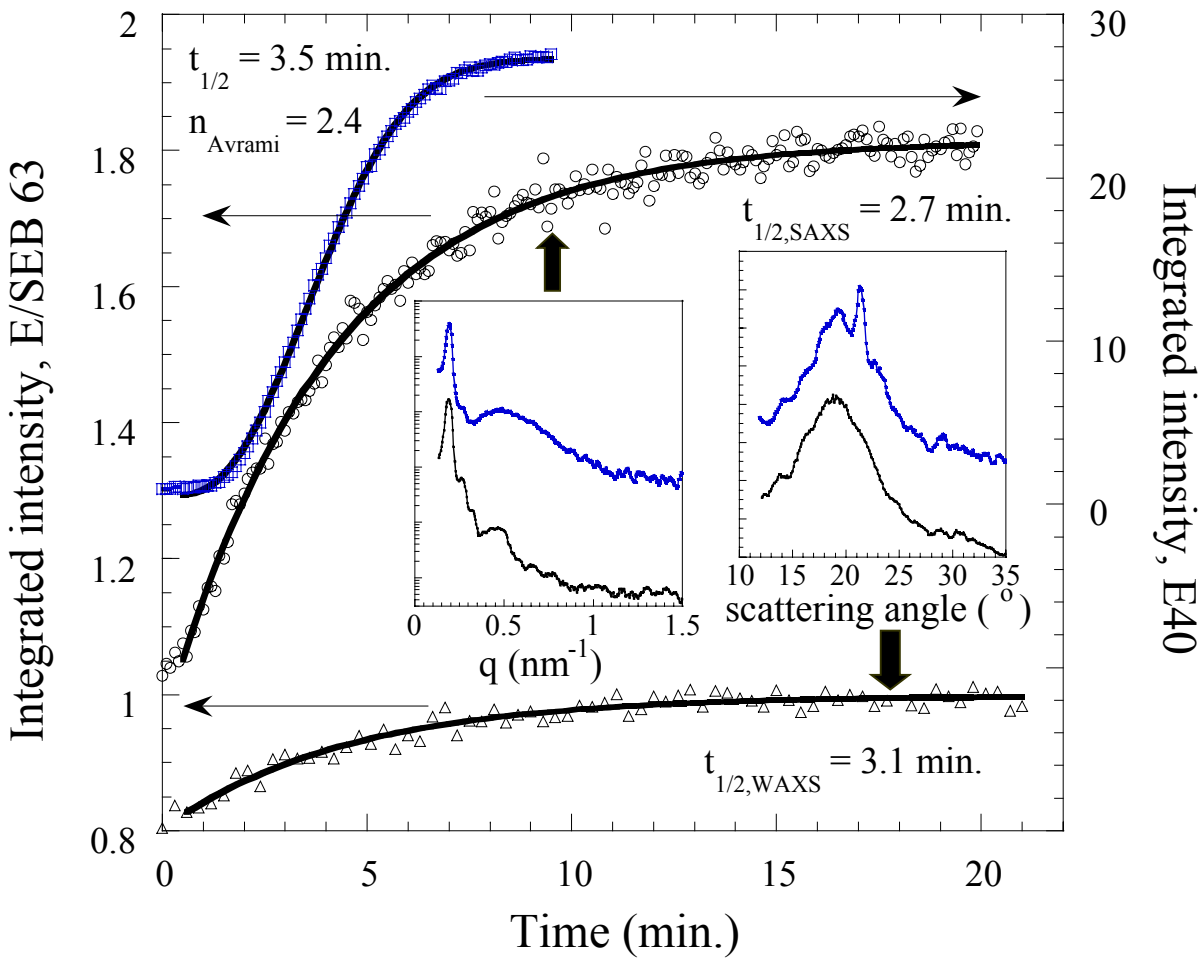

Figure 3. Time course of the integrated SAXS (middle curve) and WAXS (bottom curve) intensities for E/SEB63 crystallized at 67°C (insets show regions of integration). The SAXS intensity for the E40 "homopolymer", which shows a sigmoidal time evolution, is shown for comparison (top curve, 95°C, similar half-time).

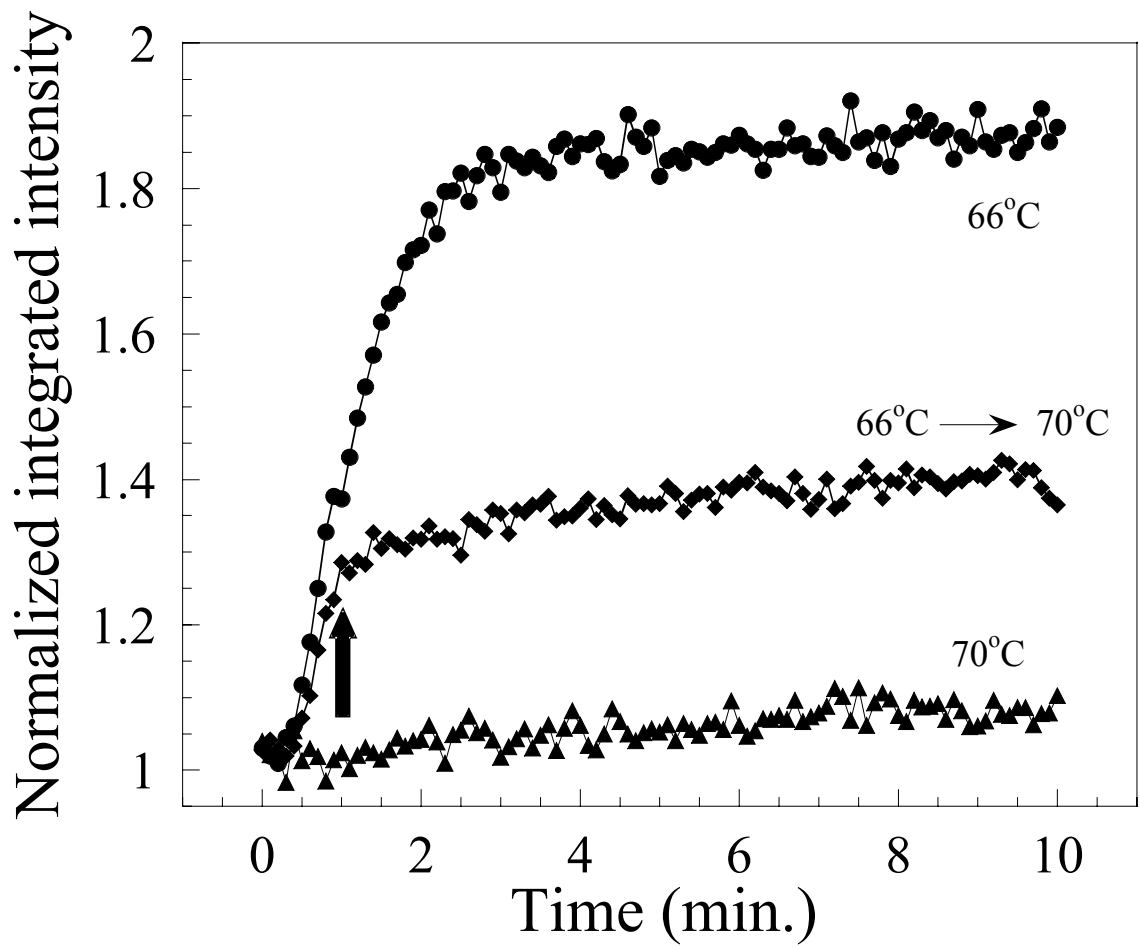

Figure 4. Time course of the integrated SAXS intensity of E/SEB63 during a jump experiment from 66 to 70°C (middle curve, jump time indicated by bold vertical arrow). Isothermal results at 66°C (top curve) and 70°C (bottom curve) are shown for comparison.

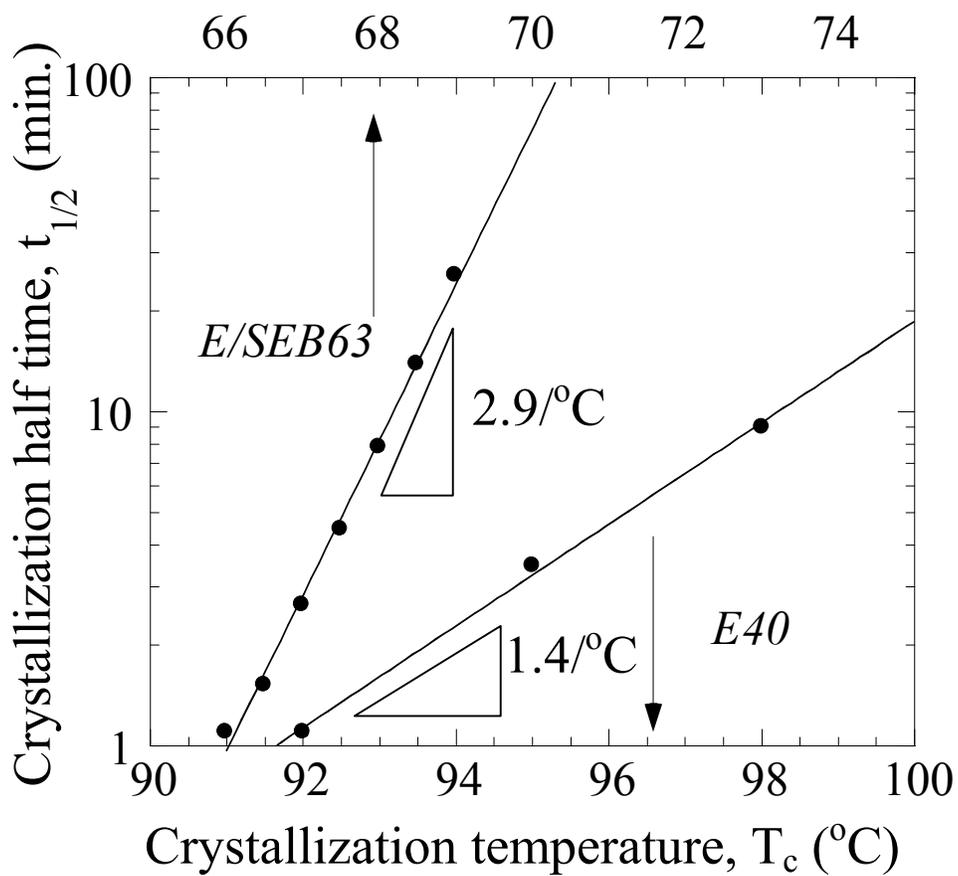

Figure 5. Dependence of isothermal crystallization half-time on temperature for E/SEB63 diblock and E40 "homopolymer".

15